\documentclass[final,3p,times,twocolumn]{elsarticle}
\usepackage{graphicx}
\usepackage{amssymb}
\usepackage{amsmath}
\usepackage{hyperref}
\usepackage{color}

\journal{Physics Letters A}

\usepackage{hyperref}
\usepackage[T1]{fontenc} 

\begin{document}

\title{Spontaneous spatial order in  two-dimensional ferromagnetic spin-orbit coupled uniform spin-1 condensate
solitons }

\author{S. K. Adhikari\footnote{s.k.adhikari@unesp.br      \\  https://professores.ift.unesp.br/sk.adhikari/ \\ \\ {\it \color{red}Preprint accepted in Physics Letters A}}}
\address{Instituto de F\'{\i}sica Te\'orica, Universidade Estadual Paulista - UNESP, 01.140-070 S\~ao Paulo, S\~ao Paulo, Brazil}
      
%%%%%%%%%%%%%%%%%%%%%%%%%%%%%%%%%%%%%%%%%%%%%%%%%%%%%%%%%%%%%%%%%%%%%%%%%%%%%%
%%%%%%%%%%                    Abstract                             %%%%%%%%%%%
%%%%%%%%%%%%%%%%%%%%%%%%%%%%%%%%%%%%%%%%%%%%%%%%%%%%%%%%%%%%%%%%%%%%%%%%%%%%%%

\date{\today}

\begin{abstract}

We demonstrate spontaneous spatial order in stripe and super-lattice solitons in a Rashba  spin-orbit
(SO) coupled   spin-1 {\it uniform} quasi-two-dimensional (quasi-2D)   ferromagnetic Bose-Einstein condensate. 
 For weak SO coupling, the solitons are of the {$(- 1,0,+ 1)$ or $(0,+ 1,+ 2)$} type with intrinsic vorticity, where the numbers in the parentheses denote angular momentum in spin components
$F_z=+1,0,-1$, respectively. For intermediate SO coupling, three types of solitons are found: (a) 
circularly-asymmetric solitons, (b) circularly-symmetric  { $(- 1,0,+ 1)$- and  (c)  $(0,+ 1,+ 2)$-type}  multi-ring solitons
maintaining the above-mentioned vortices in respective components. For large SO coupling, quasi-degenerate stripe and super-lattice   solitons  are found in addition to the circularly-asymmetric solitons.
% A stripe   soliton forms a periodic 1D  stripe pattern   in component densities.
A super-lattice soliton forms a 2D square lattice  structure in the {\it total density} as in a {\it super-solid}; in component densities it may have either (i) a 1D stripe pattern or (ii) a  2D square lattice structure.

\end{abstract}

 \maketitle

A super-solid  \cite{sprsld} is a special quantum state  where matter forms a {stable, 
spatially-ordered structure}, breaking  continuous translational invariance, as in a crystalline solid,  and  {has friction-less flow as} in a super-fluid, breaking continuous gauge invariance. In other words, it possesses qualities of 
a super-fluid and a solid,  in contradiction to the intuition that friction-less  flow  is a property exclusive to quantum fluids, e.g. Bose–Einstein condensate (BEC) \cite{bose}, and Fermi super-fluid \cite{fermi}. 
% After the  prediction of the super-solid state \cite{sprsld}, there has been great interest in finding such a state. As liquid helium has super-fluid properties, it was thought that solid helium  could host a super-solid state. 
  The search of a super-solid  helium  \cite{sprsldex}  was inconclusive \cite{sprsldex2}. 
%As a conventional solid cannot flow, it is now believed that a super-solid is a super-fluid  that can have periodic crystalline structure.
 It is possible to mimic a super-solid by an engineering with a external periodic  potential 
in a super-fluid \cite{expot}, e.g. using an optical-lattice potential.
%, or in a Rydberg-excited BEC \cite{rydberg}.
Following theoretical suggestions to create a super-solid  with dipolar \cite{losh} and finite-range \cite{finite} atomic interactions, more recently, different experimental groups confirmed its presence in a dipolar   BEC in quasi-two-dimensional (quasi-2D)  \cite{dipolar2d} and quasi-one-dimensional (quasi-1D)  \cite{dipolar1d} geometries. 
 
There have also been suggestions to create a super-solid in a spin-orbit
(SO) coupled spinor BEC \cite{solid-1/2,solid-li}. The SO coupling naturally
appears in charged electrons and controls many properties of a solid including
its crystalline or amorphous structure. Although, in a neutral atom, there
cannot be a natural SO coupling, it is possible to introduce an artificial
synthetic SO coupling by Raman lasers that coherently couple the spin-component
states in a spinor BEC \cite{thso}. Two possible SO couplings are due to
Rashba \cite{SOras} and Dresselhaus \cite{SOdre}. An equal mixture of these
SO couplings has been realized in pseudo spin-1/2 $^{87}$Rb
\cite{exptso} and $^{23}$Na \cite{na-solid} BECs containing only two spin
components $F_{z}=0,-1$ of total spin $F=1$, and in a spin-1 ferromagnetic
$^{87}$Rb BEC, containing all three spin components
$F_{z}= \pm 1, 0$ \cite{exptsp1}. A spin-1 spinor BEC
\cite{exptspinor} appears in two magnetic phases with distinct properties \cite{thspinorb}: ferromagnetic
$(a_{0}>a_{2})$ and anti-ferromagnetic $(a_{0}<a_{2})$, where
$a_{0}$ and $a_{2}$ are the scattering lengths in the total spin 0 and
2 channels, respectively. Following theoretical suggestions
\cite{solid-1/2,sinha}, recently, there has been experimental confirmation
of a periodic one-dimensional (1D) super-stripe state
with density modulation in an SO-coupled pseudo spin-1/2 spinor BEC
\cite{st2} employing an equal mixture of Rashba and Dresselhaus couplings.

In this letter, we pursue a numerical study of the spin-1 quasi-two-dimen\-sional (quasi-2D)
Rashba  SO-coupled uniform self-attractive ferromagnetic
BEC using a mean-field model. For a weakly Rashba  SO-coupled
spin-1 BEC, two types of quasi-degenerate circularly-symmetric solitons
are formed. The solitons are either of the $(- 1, 0, + 1)$ type or
of the $(0, + 1, + 2)$ type, where the numbers in parentheses represent
the angular momentum of vortices at the center of components
$F_{z} = +1, 0, -1$ \cite{kita}, the positive (negative) sign corresponds
to a vortex (an anti-vortex). For medium SO-coupling, quasi-degenerate  circularly-symmetric multi-ring
solitons are formed maintaining the above-mentioned vortices at the respective
centers in addition to circularly-asymmetric solitons without any internal vortex.
The circularly-asymmetric solitons have the lowest energy and they continue to exist for larger strengths of SO coupling.
For large SO coupling, the solitons develop spontaneous spatial order in
the form of quasi-2D crystallization onto a square lattice in total density, sharing some properties of a super-solid,
viz. Fig. 1 of Ref.~\cite{2020}, Fig. 1 of Ref.~\cite{solid-li}\footnote{Here
the equal component densities add up to make stripe in total density.}
and Fig. 1(e) of Ref.~\cite{sinha},\footnote{Here the authors do not use the name super-solid. The name super-solid
in this context was emphasized later in Ref.~\cite{solid-li}.}
 breaking
continuous translational symmetry, as in a super-solid, possessing either
1D stripe or quasi-2D crystallization in component densities and will be
termed super-lattice solitons. 
The multi-ring, viz. Figs. 3(d)-(i), and stripe, viz. Figs. 4(d)-(f), solitons only exhibit a periodic
structure in the component densities, due to a phase separation among the components without any periodic structure
in the total density. 
A 2D super-solid crystalline
phase was recently created in a $^{87}$Rb BEC dispersively coupled with
two optical cavities and illuminated by a 1D transverse pump lattice
\cite{st3}. The present SO-coupled quasi-2D super-lattice soliton without
any periodic potential in the Hamiltonian is different from this recently
observed 2D super-crystal \cite{st3} formed in the presence of a self-consistent
periodic potential generated from the interference between the transverse
pump field and the two cavity fields. The present quasi-2D super-lattice
state should serve as a laboratory model for crystallization in a solid
in a controlled environment. A similar study in a normal solid is very
limited due to impurities and lack of information about the interactions.
There have been studies of quasi-1D \cite{q1d}, quasi-2D \cite{q2d} and
three-dimensional \cite{q3d} solitons in pseudo spin-1/2 SO-coupled BEC
and no super-lattice state has been reported there.

We consider a BEC of $N$ atoms, 
{ of mass $ m$ each, free in the $x-y$ plane  and under a 
harmonic trap $V({\bf r})=    m\omega_z^2 z^2/2$ in the $z$ direction with  $\omega_z$  the} angular trap frequency.  { Integrating
out the $z$ coordinate \cite{sala}, we have the single-particle Hamiltonian
of the quasi-2D Rashba SO-coupled  BEC as} \cite{exptso,thspinorb}
%
%e1 #&#
\begin{equation}
H_{0} = -\frac{\hbar ^{2}}{2 m}\nabla _{\boldsymbol{\rho }}^{2}+
\gamma \left [ p_{y}\Sigma _{x}-p_{x} \Sigma _{y} \right ],
\label{sph}
\end{equation}
where ${\boldsymbol{\rho }}\equiv \{x,y \}$,
$\nabla _{\boldsymbol{\rho }}^{2}\equiv (\partial _{x}^{2}+\partial _{y}^{2})$,
$\partial _{x} \equiv \partial /\partial x$,
$\partial _{y} \equiv \partial /\partial y$, 
the $x$ and $y$ components of the momentum operator 
$p_{x} =-i\hbar \partial _{x}, p_{y}=-i\hbar \partial _{y}$,
$\gamma $ is the strength of Rashba SO coupling and the irreducible representation of the spin-1 spin matrices
$\Sigma _{x}$ and $\Sigma _{y}$ are
%
%e2 #&#
%e2 #&#
%e2 #&#
%e2 #&#
%e2 #&#
\begin{eqnarray}
\Sigma _{x}=\frac{1}{\sqrt{2}}
\begin{pmatrix}
0 & 1 & 0
\\
1 & 0 & 1
\\
0 & 1 & 0
\end{pmatrix}
, \quad \Sigma _{y}=\frac{i}{\sqrt{2} }
\begin{pmatrix}
0 & -1 & 0
\\
1 & 0 & -1
\\
0 & 1 & 0
\end{pmatrix}
.
\end{eqnarray}

The reduced quasi-2D coupled Gross-Pitaevskii (GP) equations
\cite{sala} of the three spin components, for the SO-coupled spin-1 spinor
BEC, are \cite{thspinorb}
%
%e3 #&#
%e4 #&#
%e5 #&#
%e6 #&#
\begin{align}\label{EQ1} 
i \partial_t &\psi_{\pm 1}({\boldsymbol 
\rho})= \left[{\cal H}+{c_2}
\left(n_{\pm 1} -n_{\mp 1} +n_0\right)  \right] \psi_{\pm 1}({\boldsymbol 
\rho})\nonumber \\
+&\left\{c_2 \psi_0^2({\boldsymbol \rho})\psi_{\mp 1}^*({\boldsymbol \rho})\right\} % \nonumber \\
-i {\widetilde \gamma} (\partial_y \pm i \partial_x)  \psi_{0}  ({\boldsymbol \rho})  \, , 
\\ \label{EQ2}
i \partial_t&\psi_0({\boldsymbol \rho})=\left[ {\cal H}+{c_2}
\left(n_{+ 1}+n_{- 1}\right) \right] \psi_{0}({\boldsymbol \rho})+\big \{2c_2 \psi_{+1}({\boldsymbol \rho}) \nonumber \\
\times &\psi_{-1}({\boldsymbol \rho})\psi_{0}^* ({\boldsymbol \rho})\big\}   
{-i{\widetilde \gamma} [-i  \partial_x \phi_{-1}  ({\boldsymbol \rho})  +  \partial_y \phi_{+1} ({\boldsymbol \rho}) ]}  
  \, , \\
{\cal H}=&-\frac{1}{2}\nabla^2_{\boldsymbol \rho}+ c_0 n,    \\
c_0 =&\frac{2{N}\sqrt{2\pi}(a_0+2a_2)}{3}, \quad c_2 
= \frac{2{N}\sqrt{2\pi}(a_2-a_0)}{3}, \label{EQ4}
\end{align}
where $\widetilde{\gamma }= \gamma /\sqrt{2} $,
$\partial _{t} \equiv \partial /\partial t$,
$\phi _{\pm } ({\boldsymbol{\rho }}) =\psi _{+1} ({\boldsymbol{\rho }})
\pm \psi _{-1}({\boldsymbol{\rho }})$,
$n_{j} = |\psi _{j}|^{2}, j=\pm 1,0$ are the densities of spin components
$F_{z}= \pm 1 , 0$, the total density
$n ({\boldsymbol{\rho }})= \sum _{j} n_{j}({\boldsymbol{\rho }})$, and the
asterisk denotes complex conjugate. All quantities in Eqs.~{(\ref{EQ1})}-{(\ref{EQ4})}
are dimensionless, as we express lengths ($a_{0},a_{2},x,y,z$) in units
of oscillator length $l_{0}\equiv \sqrt{\hbar / m\omega _{z}}$, density
in units of $l_{0}^{-2}$, time in units of $\omega _{z}^{-1}$, and energy
in units of $\hbar \omega _{z}$. The normalization condition is
$ {\textstyle \int }n({\boldsymbol{\rho }})\, d{\boldsymbol{\rho }}=1$. A spin-1
spinor BEC appears in two magnetic phases: ferromagnetic ($c_{2}<0$) and
anti-ferromagnetic ($c_{2}>0$). The stationary solutions are governed by
the time-independent version of Eqs.~{(\ref{EQ1})}-{(\ref{EQ2})} with the energy
functional
%
%e7 #&#
\begin{align}
\label{energy}
E[\psi ] &= \frac{1}{2} {\textstyle \int } d{\boldsymbol{\rho }} \big [ {
\textstyle \sum }_{j} |\nabla _{\boldsymbol{\rho }}\psi _{j}|^{2}-2\mu
n+
\rho ^{2}n+c_{0}n^{2%
}
\nonumber \\
& + c_{2}\big \{n_{+1}^{2} +n_{-1}^{2} +2(n_{+1}n_{0}+n_{-1}n_{0}-n_{+1}n_{-1}
\nonumber
\\
&\times +\psi _{-1}^{*}\psi _{0}^{2}\psi _{+1}^{*} + \psi _{-1}\psi _{0}^{*2}
\psi _{+1}) \big \}
\nonumber \\
&-2i\widetilde{\gamma }\big \{ \psi _{0}^{*}
\partial _{y} \phi _{+}%
+  \phi _{+}^{*} \partial _{y} \psi _{0} -i \psi _{0}^{*}
\partial _{x} \phi _{-} +i \phi _{-}^{*}\partial _{x} \psi _{0}
\big \} \big ] +\mu,\qquad
\end{align}
where $\mu$ is the chemical potential. However, the numerical value of energy is independent of the chemical potential.

\begin{figure}[!t] 
\centering
 \includegraphics[width=\linewidth]{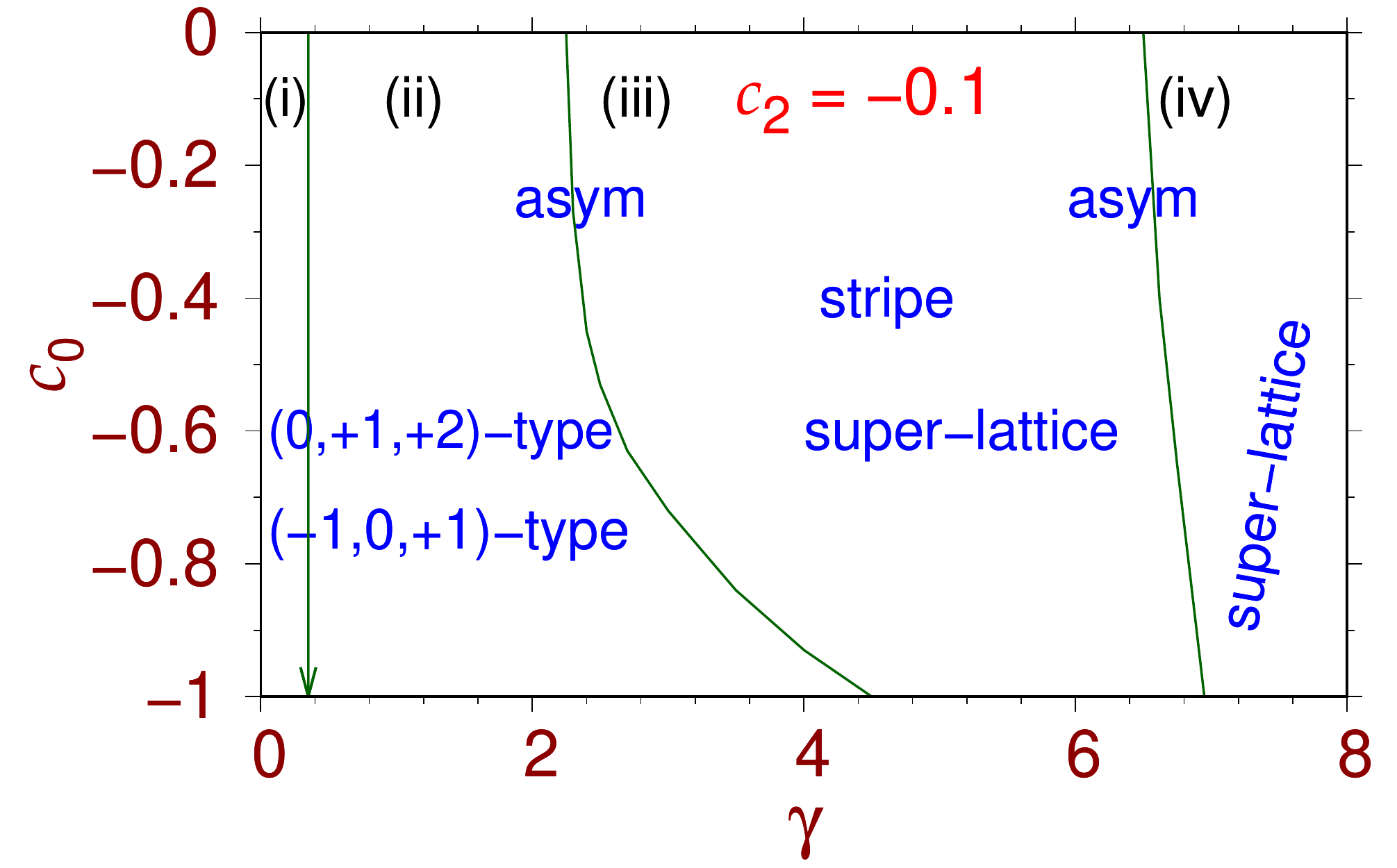}
\caption{ The $c_0$ versus $\gamma$  phase plot for Rashba  SO coupling
 in different regions of parameter space: (i,ii){$(0,+ 1, + 2)$ and $(- 1,0, + 1)$-type} solitons;
(ii,iii,iv) circularly asymmetric soliton; (iii) stripe and super-lattice solitons;  (iv) super-lattice solitons.      Results in all Figs. are in dimensionless units. }
\label{fig1}

\end{figure}

\begin{figure}[!t] 
\centering
\includegraphics[width=.325\linewidth]{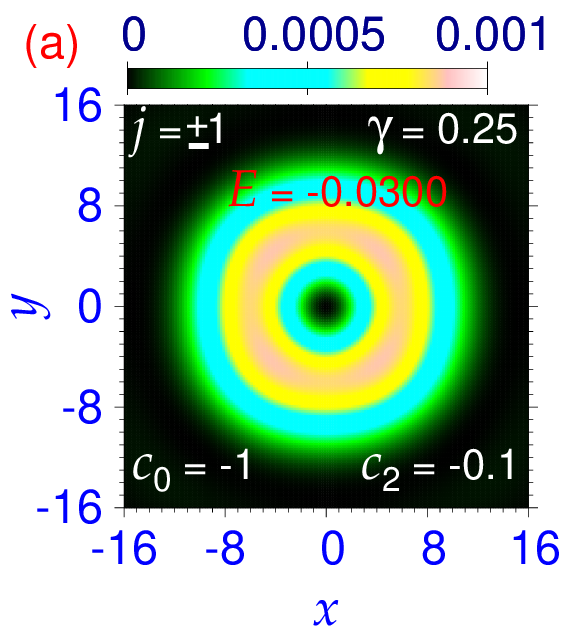} 
\includegraphics[width=.325\linewidth]{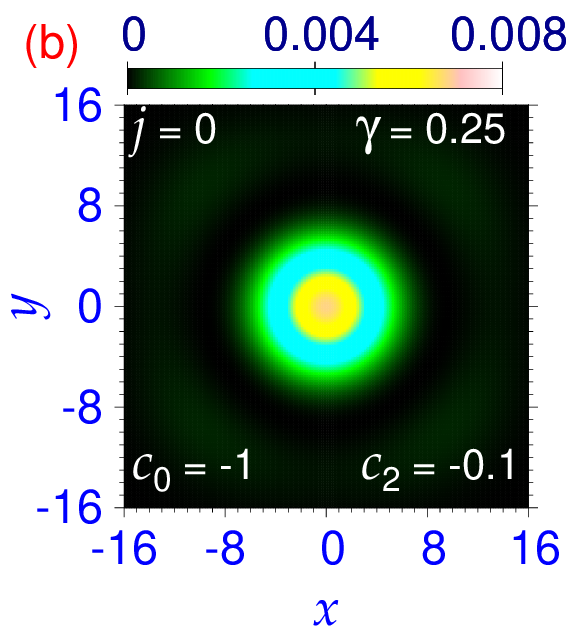}
\includegraphics[width=.325\linewidth]{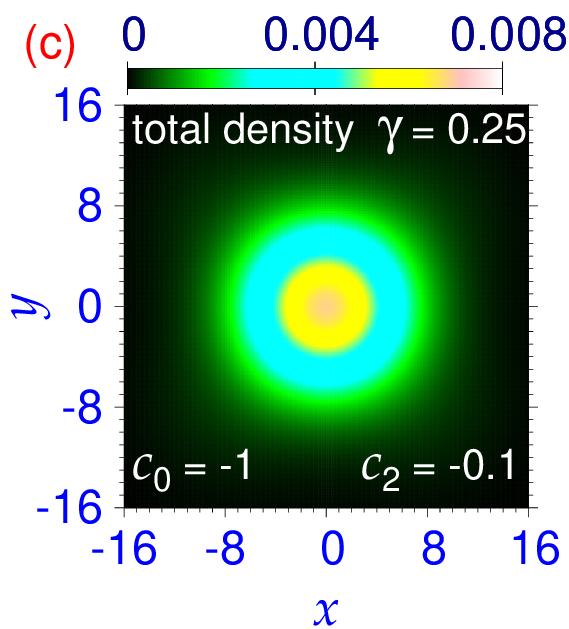}

 \includegraphics[width=.325\linewidth]{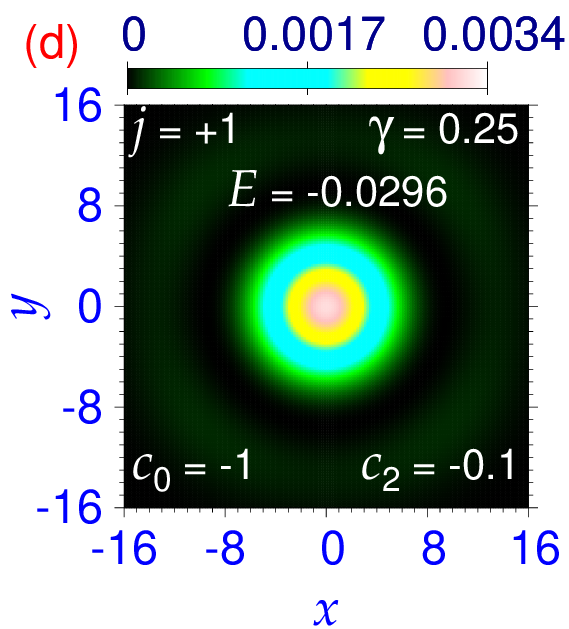} 
\includegraphics[width=.325\linewidth]{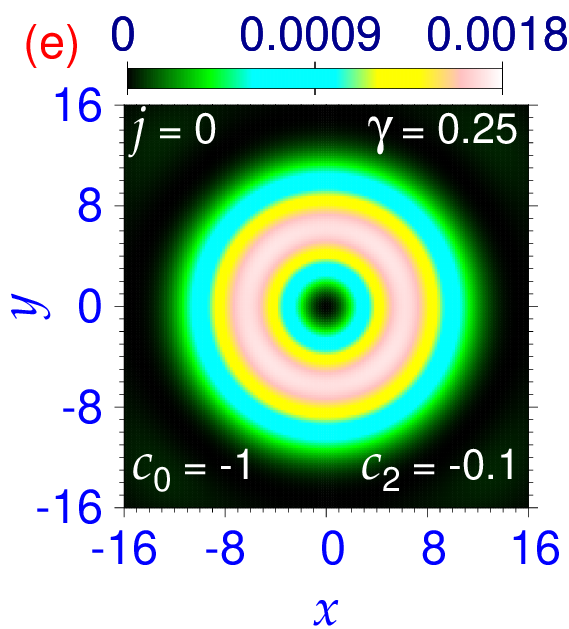}
 \includegraphics[width=.325\linewidth]{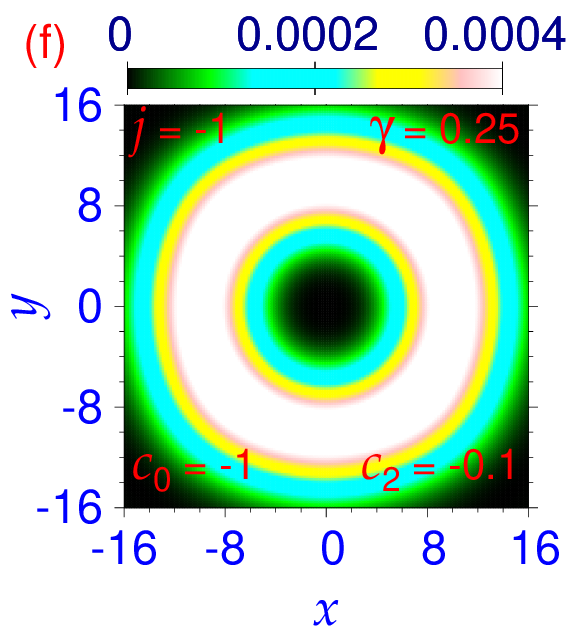}

 \includegraphics[width=.23\linewidth]{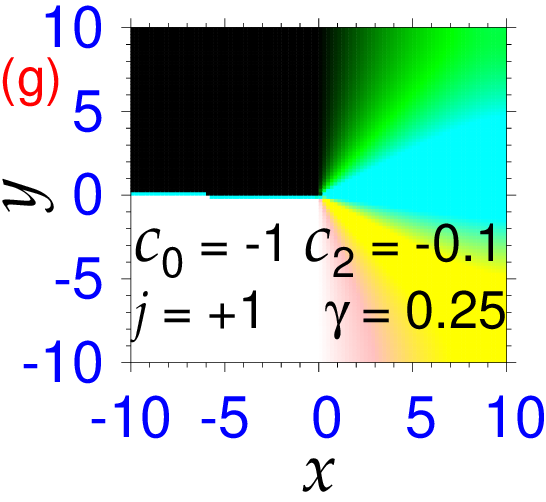} 
\includegraphics[width=.23\linewidth]{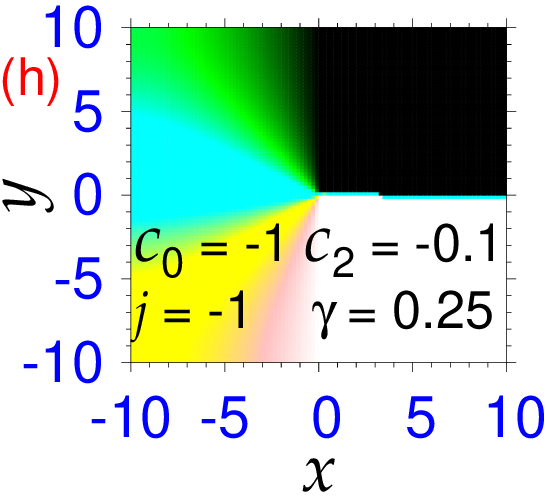}
\includegraphics[width=.23\linewidth]{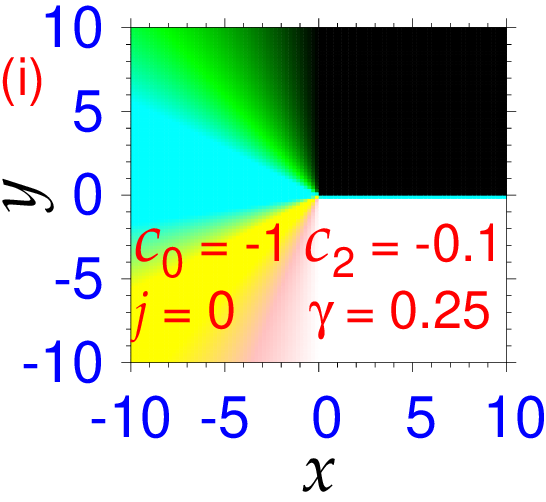}
 \includegraphics[width=.26\linewidth]{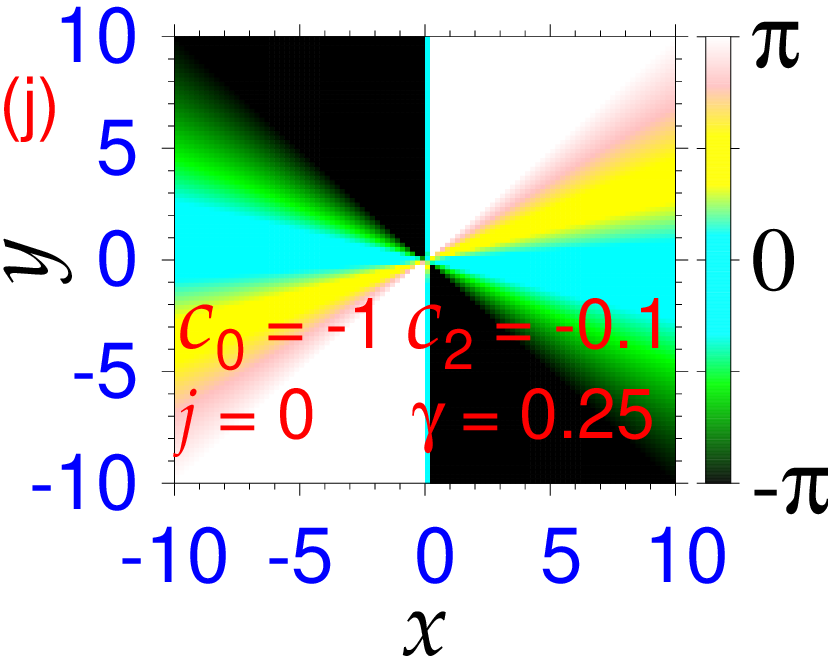}

\caption{Contour plot of density $n_j(\boldsymbol  \rho)$ of components (a) $j= \pm 1$, (b) $j=0$ and  (c) total density $n(\boldsymbol  \rho)$
 of a {$(- 1,0,  + 1)$-type} Rashba  SO-coupled   soliton  for $c_0=-1,c_2=-0.1, \gamma=0.25$;  the same  of 
components (d) $j=+1,$ (e)  $j=0$, (f) $j= -1$ of a {$(0,+ 1,+ 2)$}-type  soliton; the phase of wave-function components
 (g) $j=+1$ and (h) $j=-1$  of  the Rashba SO-coupled  
$(-1,0,+1)$-type BEC soliton;    the phase of (i) $j=0$ and (j) $j=-1$ components of  the Rashba SO-coupled  
$(0,+1,+2)$-type BEC soliton. 
The energies are displayed in the $j=+1$ component in Figs. \ref{fig2}-\ref{fig5}.}
\label{fig2}

\end{figure}

We solve Eqs.~{(\ref{EQ1})} and {(\ref{EQ2})} numerically propagating these
in time by the split-time-step Crank-Nicolson discretization scheme
\cite{bec2009} employing a space step of 0.1 and a time step of 0.001 and
0.00025, respectively, for imaginary- and real-time propagation. Imaginary-time
propagation is used to find the lowest-energy soliton in different parameter
domains. Real-time propagation is used to test the dynamical stability
of the solution. Magnetization is not a good quantum number and
evolves freely during time propagation to attain a final converged value.

To simulate a self-attractive ($c_{0}<0$) ferromagnetic ($c_{2}<0$) SO-coupled
spinor BEC for different strengths of SO coupling $\gamma $, we consider
a system with $c_{2}=-0.1$ and vary $c_{0}$ and $\gamma $ in this study.
We consider $c_{0}$ in the range $0>c_{0}>-1$, as for $ c_{0}<-1$, because
of increased attraction, the soliton shrinks and its size becomes too small
for a useful calculation. In {Fig.~\ref{fig1}} we illustrate the formation
of different types of solitons for Rashba  SO coupling via
the phase plot of $c_{0}$ versus $\gamma $ for $c_{2}=-0.1$. Solitons of
type $(- 1,0,+ 1)$ and $(0,+ 1,+ 2)$ are formed in region (i) as
well as (ii); in the second they evolve into multi-ring solitons. Circularly-asymmetric
solitons are formed in regions (ii), (iii), and (iv). Super-lattice solitons
are formed in regions (iii) and (iv). Stripe solitons are formed in region
(iii).

\begin{figure}[!t] 
\centering
\includegraphics[width=.325\linewidth]{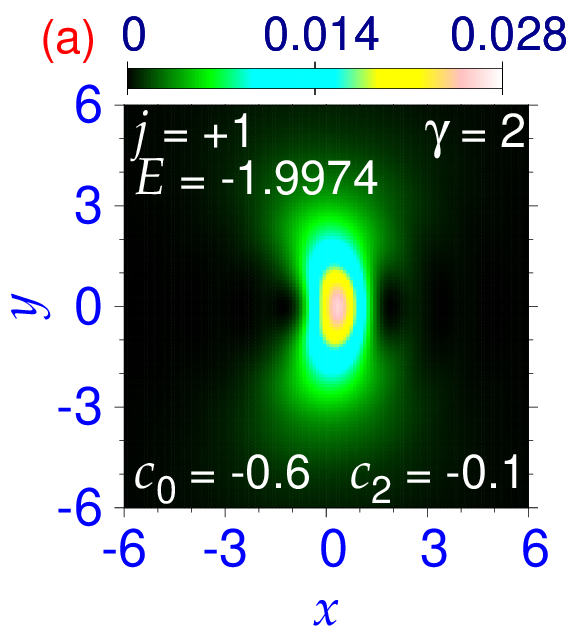} 
\includegraphics[width=.325\linewidth]{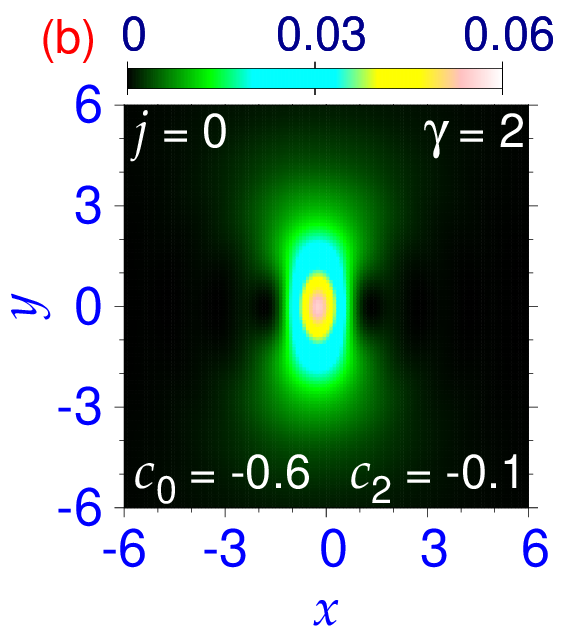}
\includegraphics[width=.325\linewidth]{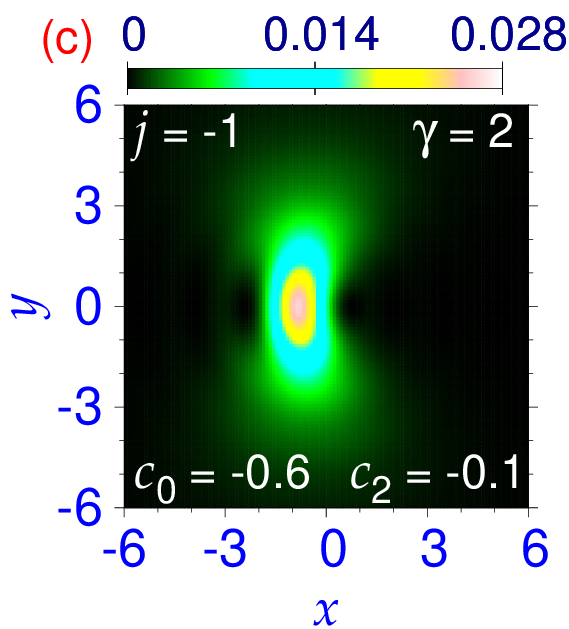}

 \includegraphics[width=.325\linewidth]{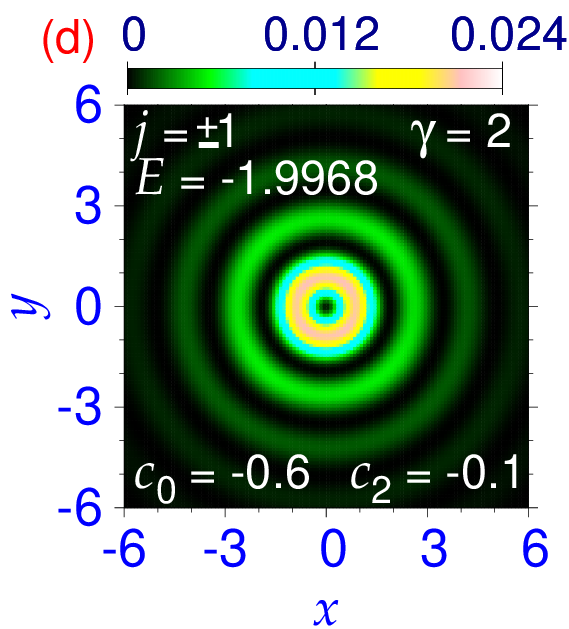} 
\includegraphics[width=.325\linewidth]{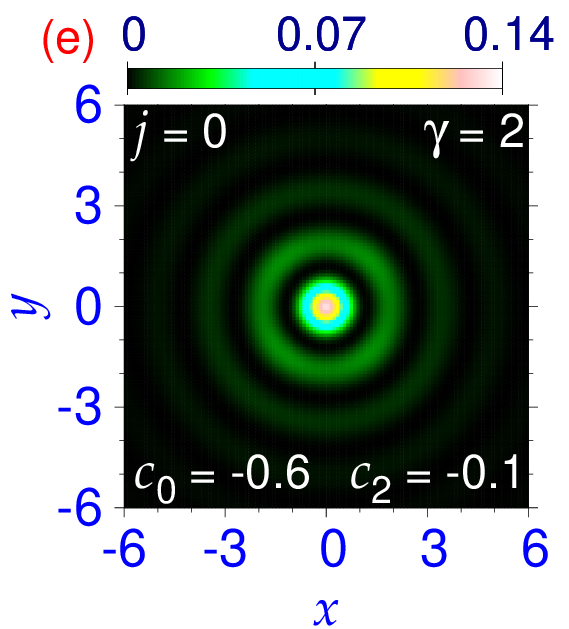}
\includegraphics[width=.325\linewidth]{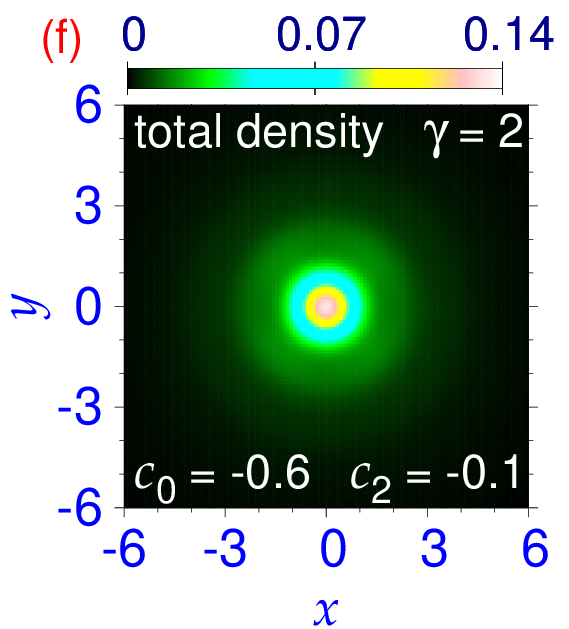}

 \includegraphics[width=.325\linewidth]{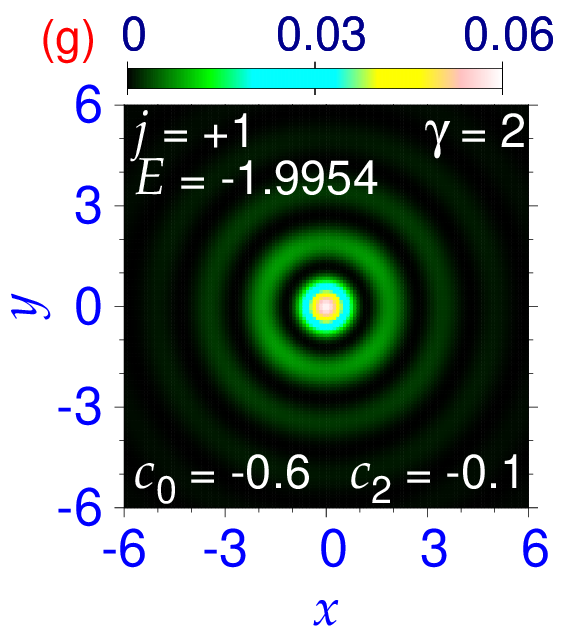} 
\includegraphics[width=.325\linewidth]{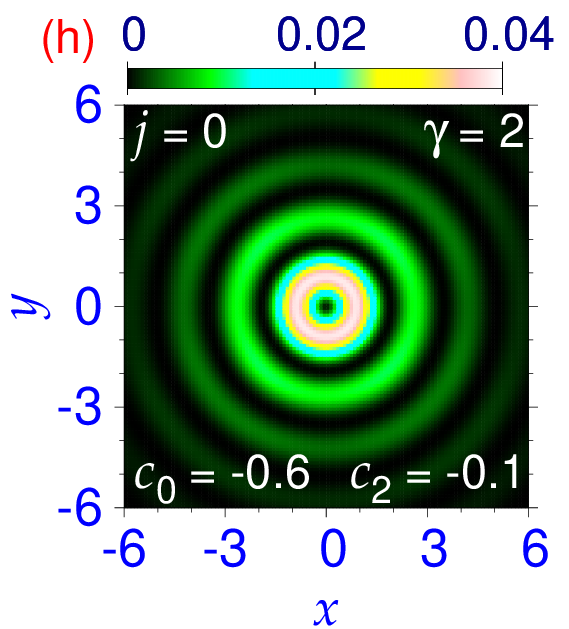}
\includegraphics[width=.325\linewidth]{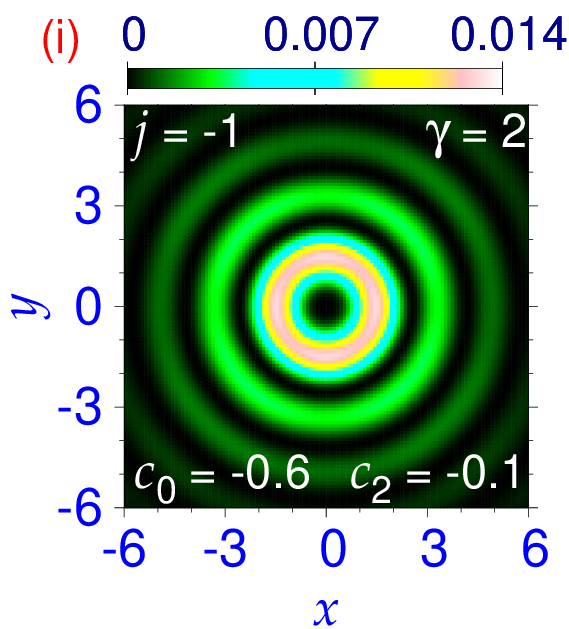}
\

\caption{ Contour plot of density $n_j(\boldsymbol \rho)$ of wave-function components (a) $j=+1$, (b) $j=0$ and  (c) $j=-1$ of
the asymmetric Rashba  SO-coupled 
soliton with $c_0=-0.6, c_2=-0.1, \gamma=2$; the same of the  
 {$(- 1,0,+ 1)$}-type  soliton of components (d) $j=\pm 1$, (e) $j=0$, and 
(f) the total density $n(\boldsymbol \rho)$; 
the same  of the  {$(0,+ 1, + 2)$}-type  soliton  of components (g) $j=+1$, (h) $j=0$ and  (i) $j=-1$.    }
\label{fig3}

\end{figure}

First we consider a small SO-coupling strength ($\gamma =0.25$), where
two quasi-degenerate vector solitons of type $(- 1,0,+ 1)$ and
$(0,+ 1,+ 2)$ are formed for Rashba  SO coupling obtained by imaginary-time propagation using initial
localized functions with the appropriate vortices imprinted in the respective
components. In {Figs.~\ref{fig2}}(a)-(c) we display the contour plot of density
of components (a) $j=\pm 1$ and (b) $j=0$ and (c) total density of a
$(- 1,0,+ 1)$-type soliton for parameters
$c_{0}=-1,c_{2}=-0.1,\gamma =0.25$. In {Figs.~\ref{fig2}}(d)-(f) we show
the same of components (a) $j= + 1$ and (b) $j=0$ and (c) $j=-1$ of a
$(0,+ 1, + 2)$-type soliton for the same parameters. The energy of
the $(- 1,0,+ 1)$- and $(0,+ 1, + 2)$-type solitons in {Fig.~\ref{fig2}} are $E=-0.0300$ and $E=-0.0296$ respectively. In {Figs.~\ref{fig2}}(g)-(h)
we display the contour plot of the phase of wave-function components
$j=\pm 1$ of the $(-1,0,+1)$-type Rashba SO-coupled soliton of {Fig.~\ref{fig2}}(a) showing a phase drop of $\mp 2\pi $ under a complete rotation,
indicating angular momenta of $\mp 1$ in these components. In {Figs.~\ref{fig2}}(i)-(j) we display the same of components $j=0, -1$ of the
$(0,+1,+2)$-type Rashba SO-coupled soliton of {Figs.~\ref{fig2}}(e)-(f) showing
a phase drop of $2\pi $ and $4\pi $ under a complete rotation, indicating
angular momenta of $+1$ and $+2$ in these components.   The phases are consistent with the vortex (anti-vortex) structure
of the $(- 1,0,+ 1)$ and $(0,+ 1,+ 2)$-type solitons.

\begin{figure}[!t] 
\centering
\includegraphics[width=.325\linewidth]{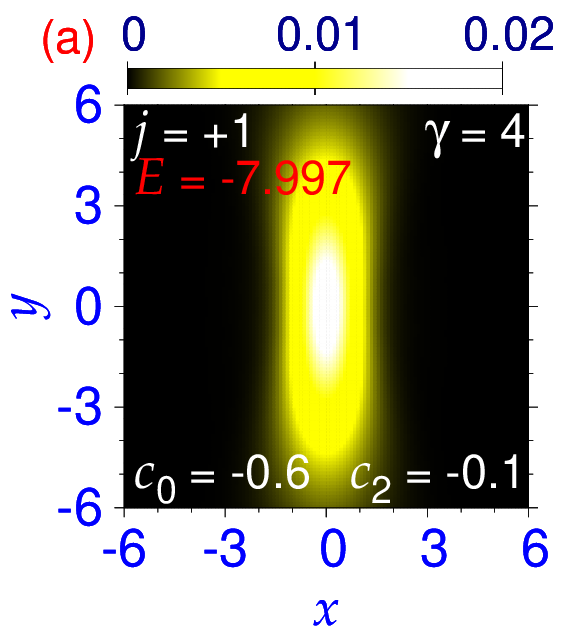}
\includegraphics[width=.325\linewidth]{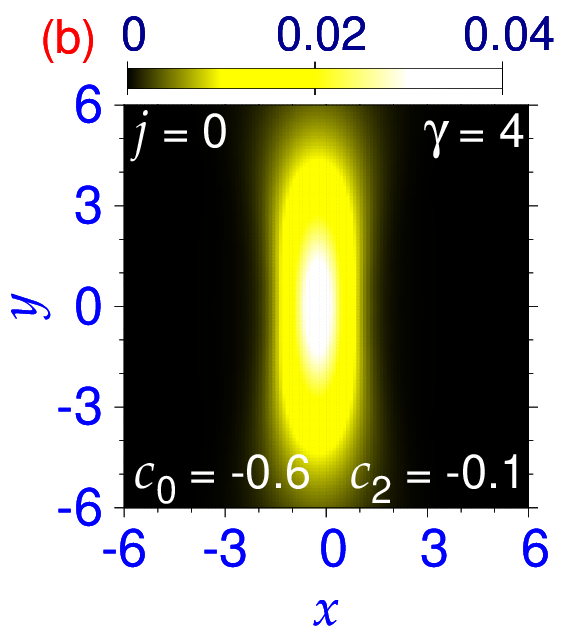}
\includegraphics[width=.325\linewidth]{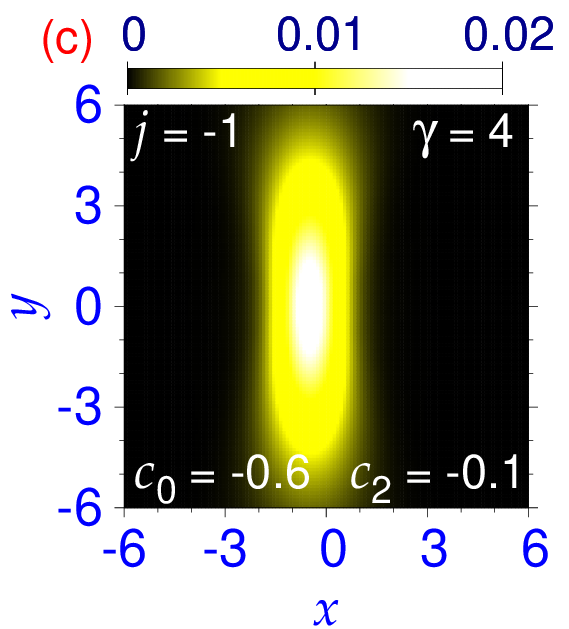}

 \includegraphics[width=.325\linewidth]{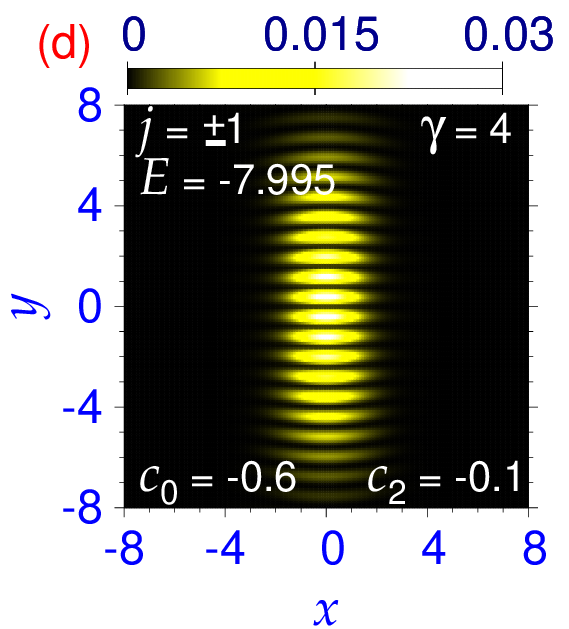} 
\includegraphics[width=.325\linewidth]{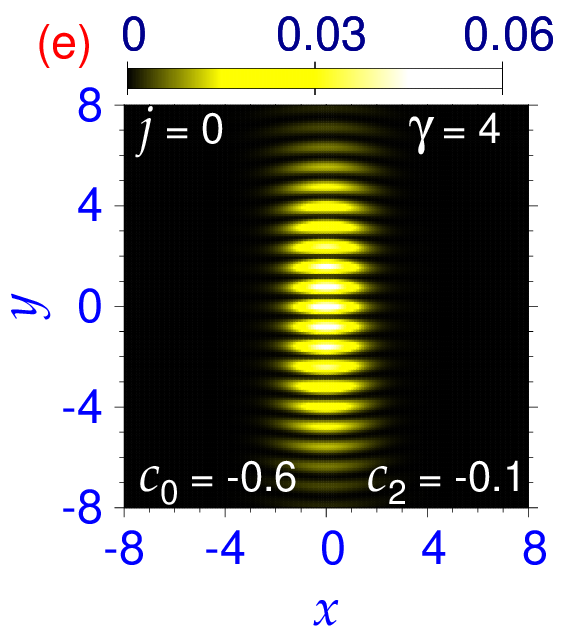}
\includegraphics[width=.325\linewidth]{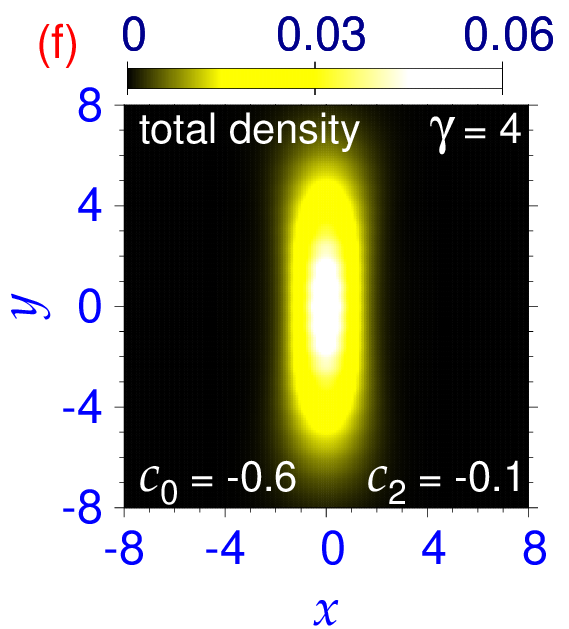}

\includegraphics[width=.325\linewidth]{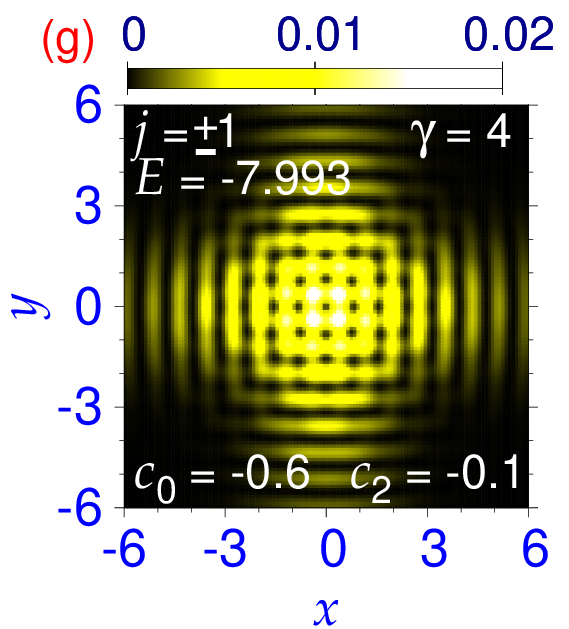} 
\includegraphics[width=.325\linewidth]{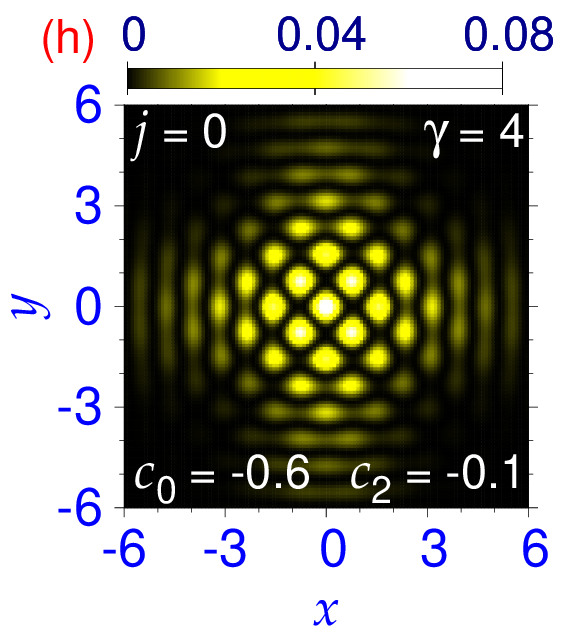}
\includegraphics[width=.325\linewidth]{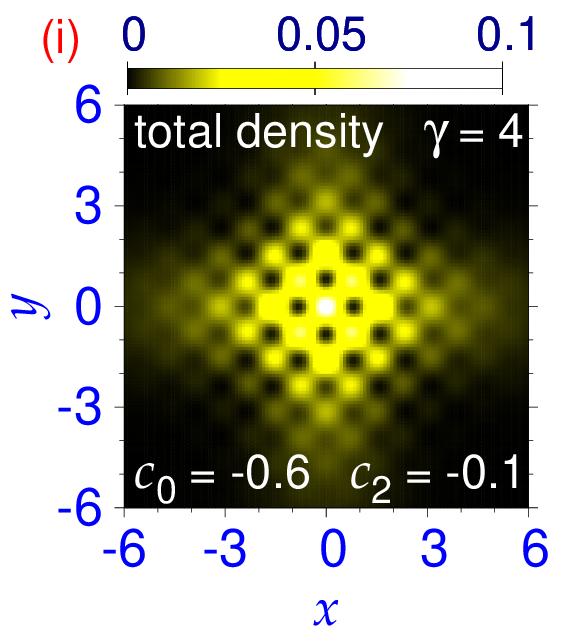}

\caption{ Contour plot of density $n_j(\boldsymbol \rho)$ of components (a) $j=+1$, (b) $j=0$ and  (c) $j=-1$ of
the asymmetric Rashba  SO-coupled 
soliton with $c_0=-0.6, c_2=-0.1, \gamma=4$;  the same of components (d) $j=\pm 1$, (e) $j=0$, and 
(f) the total density $n(\boldsymbol \rho)$ of  the  
 stripe soliton; 
(g)-(i) the same  of the  super-lattice  soliton.    }

\label{fig4}

\end{figure}

\begin{figure}[!t] 
\centering
\

\includegraphics[width=.325\linewidth]{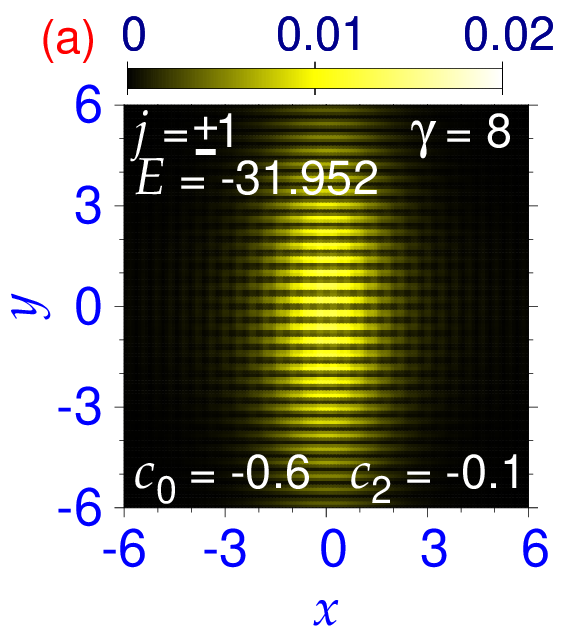} 
\includegraphics[width=.325\linewidth]{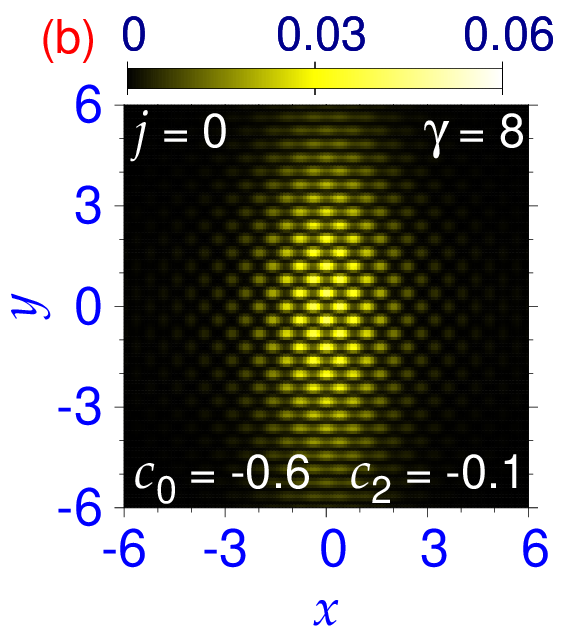}
\includegraphics[width=.325\linewidth]{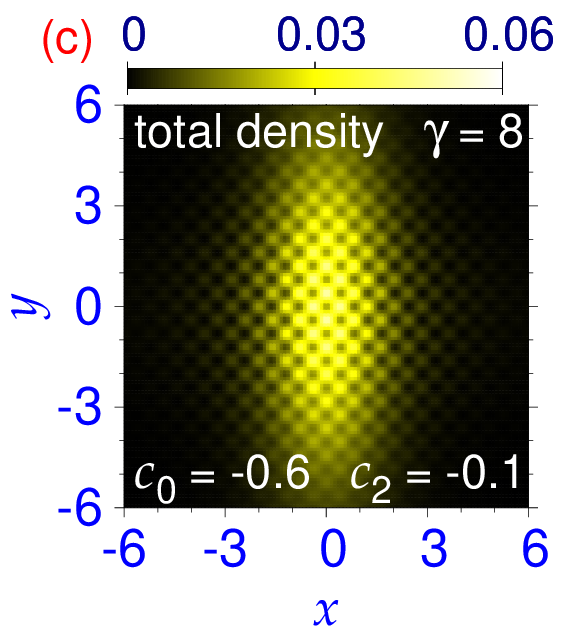}

 \includegraphics[width=.325\linewidth]{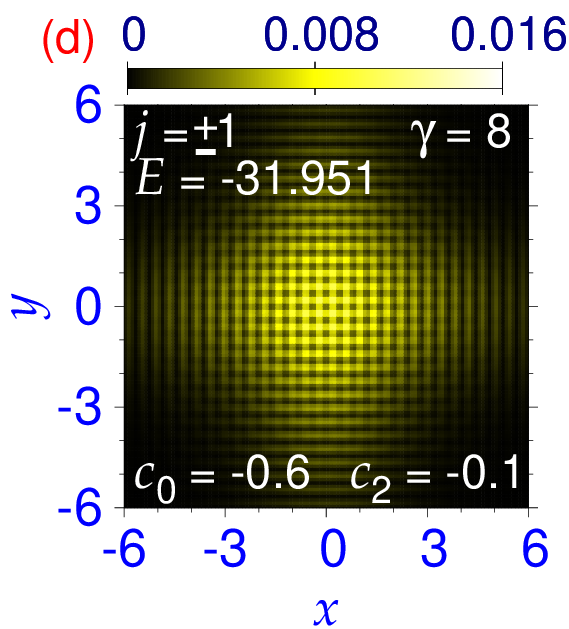} 
\includegraphics[width=.325\linewidth]{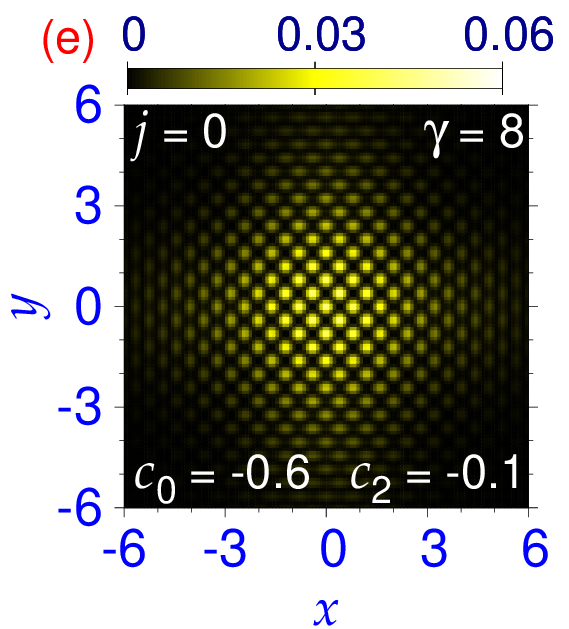}
\includegraphics[width=.325\linewidth]{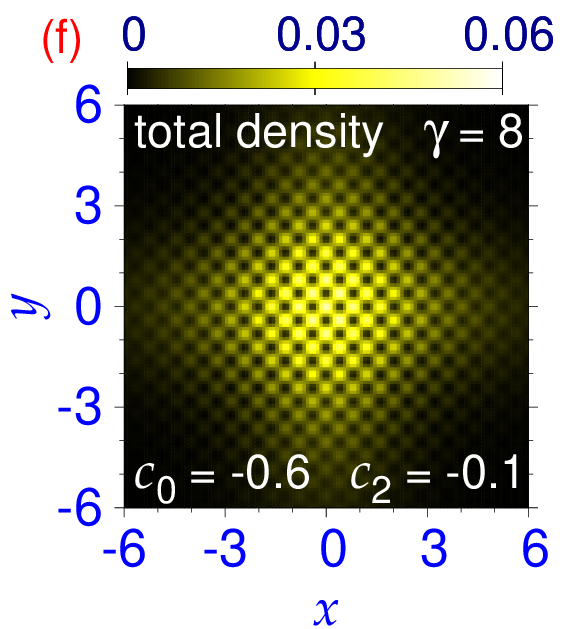}

\caption{ Contour plot of density $n_j(\boldsymbol \rho)$ of components (a) $j=\pm 1$, (b) $j=0$ and  (c)
the total density $n(\boldsymbol \rho)$ of
the super-lattice Rashba  SO-coupled 
soliton with stripe in components $j=\pm 1$ for  $c_0=-0.6, c_2=-0.1, \gamma=8$; (d)-(f) the same  of the  super-lattice  soliton with 2D square lattice formation in components $j=\pm 1$.    }
\label{fig5}

\end{figure}

Three types of quasi-degenerate vector solitons are found as $\gamma $ is increased. In
addition to the circularly-symmetric solitons an asymmetric soliton appears
as the lowest-energy state, viz. {Figs.~\ref{fig3}}(a)-(c) for
$c_{0} =-0.6, c_{2} =- 0.1$ and $\gamma =2$, illustrating the
$j=+1,0,-1$ components, respectively. The $(- 1, 0, + 1)$- and
$(0,+ 1, + 2 )$-type solitons develop concentric radially-periodic
rings around a central core maintaining the appropriate vortices in the
respective components and become $(- 1, 0, + 1)$- and
$(0,+ 1, + 2 )$-type multi-ring solitons, as shown in {Figs.~\ref{fig3}}(d)-(f) and (g)-(i), respectively. The phases of the wave-function
components of the multi-ring solitons (not shown here) are identical to
the same of the $(- 1,0,+ 1)$- and ($0,+ 1,+ 2$)-type solitons
of {Figs.~\ref{fig2}}(g)-(j) reflecting the same vortex (anti-vortex) structure
at the center of the respective components for Rashba  SO
coupling. The total density in all these {solitons have a} smooth distribution
of matter as shown in  Fig. \ref{fig3}(f) for the $(-1,0,+1)$-type multi-ring soliton. Of these three types of quasi-degenerate solitons, the circularly-asymmetric state of energy $E=-1.9974$ is the lowest-energy state. 
The first excited state is the $(-1,0,+1)$-type state with energy 
$E=-1.9968$; the next state is the $(0,+1,+2)$-type state with energy
$E=-1.9954$.
The increase of $\gamma $ from {Fig.~\ref{fig2}} to {Fig.~\ref{fig3}} has increased
the binding, and hence aids in forming soliton. The change in binding due
to the change of $c_{0}$ from {Fig.~\ref{fig2}} to {Fig.~\ref{fig3}} is negligible
in this scale. 
Multi-ring solitons were also investigated in a quasi-2D
pseudo spin-1/2 SO-coupled BEC trapped in a radially-periodic potential
\cite{radper} which creates the multi-ring modulation in component densities due to the presence of the external radially-periodic trap.
However, the present radial modulation in density without any external
trap is a consequence of the Rashba SO coupling.
 
The single-particle Hamiltonian {(\ref{sph})} should have solutions of the
plane wave form
$\exp (\pm i\alpha \gamma x)\otimes \exp (\pm i\beta \gamma y)$, where
$\alpha $ and $\beta $ are constants. In the presence of interaction ($c_{0},c_{2}
\ne 0$), the solution will be a superposition of such plane wave solutions
leading to a periodic variation of density in the form
$\sin ^{2}(\alpha \gamma x)$, $\cos ^{2}(\beta \gamma y)$,
$\sin ^{2}(\alpha \gamma x)\sin ^{2}(\beta \gamma y)$ etc. appropriate
for stripe or lattice solitons. This has been demonstrated in details for
quasi-1D solitons \cite{quasi-1d1}, the same of the present quasi-2D solitons
will be the subject of a future investigation. The period of the lattice
or stripe increases as $\gamma $ is reduced. For small $\gamma $, the size
of the soliton is smaller than this period and the periodic pattern in
density is not possible.

As $\gamma $ is increased, the $(- 1,0,+ 1)$ and
$(0,+ 1, + 2)$-type multi-ring solitons are no longer the lower-energy
states and become excited states. Two new types of solitons \textit{without}
any vortex at the center of the components: stripe and super-lattice solitons
with periodic distribution of matter in $x$ and/or $y$ directions become
the states with lower energy. However, the circularly-asymmetric soliton
continues as the lowest-energy ground state. The density profiles of the
circularly-asymmetric, stripe and super-lattice solitons are shown in {Figs.~\ref{fig4}}(a)-(c), (d)-(f), and (g)-(i), respectively, for
$c_{0}=-0.6, c_{2}=-0.1, \gamma =4$. The stripe solitons are obtained by
imaginary-time propagation using localized initial functions with appropriate
stripes imprinted in the form $\cos (\gamma y)$ and
$\sin (\gamma y)$ in components $j=\pm 1$ and 0, respectively; the super-lattice solitons are obtained using the converged
solution of {Figs.~\ref{fig3}}(d)-(f) as the initial state. The stripe soliton
of {Figs.~\ref{fig4}}(d)-(f) has 1D stripes in component densities but the
total density has a smooth distribution of matter. The super-lattice soliton
of {Figs.~\ref{fig4}}(g)-(i) has a 2D square lattice pattern in both component
densities as well as total density \cite{2020}. The
present super-lattice soliton is a consequence of the SO coupling and breaks
\textit{continuous} translational symmetry
\cite{sprsld}. For the same set of parameters, the asymmetric ground, stripe,
super-lattice states are almost degenerate with respective energies
$E=-7.997,-7.995, -7.993$.

%Although, there is a stripe pattern in density in this case,
%the positions of maxima in components $j=\pm 1$ coincide with the minima in component $j=0$ due to a phase separation among the components, thus resulting in a total density without modulation.
% In Fig. \ref{fig4x}
%the super-lattice soliton for the same set of parameters 
%is displayed through a contour plot of density of components (d) $j=\pm  1$, (e) $j=0$, and (f)  total density \cite{2020}.  
%The distribution of matter on a 2D square lattice is prominent in the total density plot of Fig.  \ref{fig4x}(f).

 As $\gamma $ is increased further $(\gamma =8)$, the asymmetric soliton,
viz. \ref{fig4}(a)-(c), continue to exist as {the} ground state with
density profile very similar to those for $\gamma =4$ (not shown here).
The stripe solitons become super-lattice solitons as displayed in {Figs.~\ref{fig5}}(a)-(c) for $c_{0}=-0.6, c_{2}=-0.1, \gamma =8$, whereas the
super-lattice solitons with 2D square lattice structure in components continue
to exist. The components $j=\pm 1$ of the former soliton have stripes in
density; whereas the density of the $j=0$ component and the total density
develop a lattice structure. The component densities as well as the total
density of a super-lattice soliton  exhibit a periodic pattern on a
2D square lattice, viz. {Fig.~\ref{fig5}} for densities of components (d)
$j=\pm 1$, (e) $j=0$, and (f) the total density. The asymmetric ground-state (not shown here),
and the two types of super-lattice solitons have energies
$E=-31.954$, $E= -31.952$ and  $-31.951$,  respectively.

\begin{figure}[!t] 
\centering
\includegraphics[width=.325\linewidth]{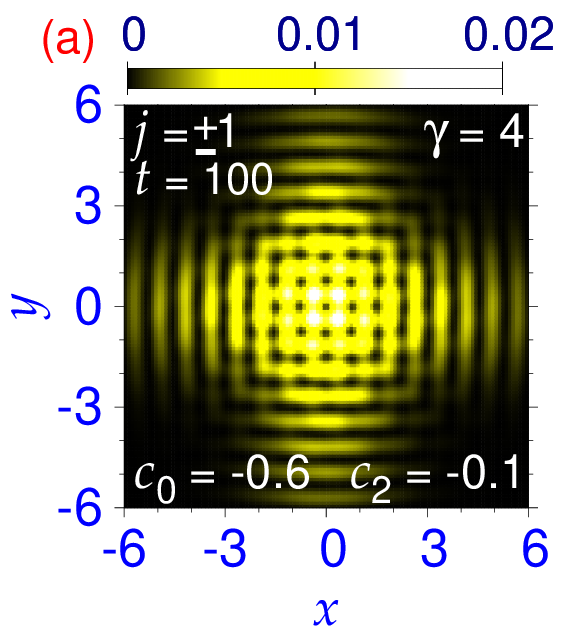} 
\includegraphics[width=.325\linewidth]{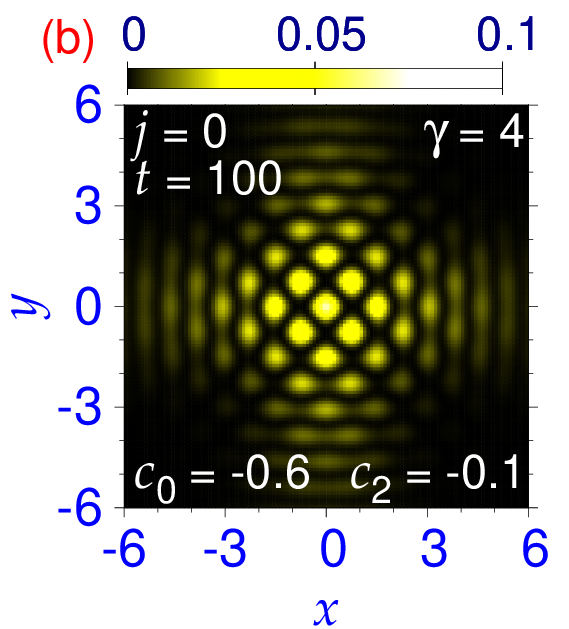}
\includegraphics[width=.325\linewidth]{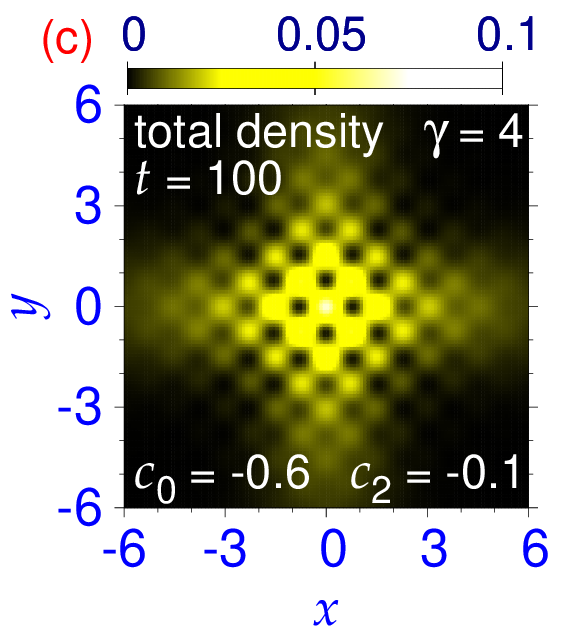}
 
\
\caption{  Contour plot of density of the super-lattice  soliton, of
Fig. 4(g)-(i), of components (a) $ j = \pm 1,$ (b) $j = 0$ and (c) total density, after real-
time propagation at time $t = 100$, using the converged imaginary-time wave function as the initial state.     }
\label{fig6}

\end{figure}

The solitons considered here are dynamically stable. As an example, to
demonstrate the dynamical stability, we consider {the} super-lattice
soliton of {Figs.~\ref{fig4}}(d)-(f) and subject the corresponding imaginary-time
wave functions to real-time propagation during 100 units of time. The resultant
real-time densities at time $t=100$ of the super-lattice soliton are displayed
in {Figs.~\ref{fig6}} for components (a) $j=\pm 1$, (b) $j=0$ and (c) the
total density. Although the root-mean-square sizes and energy were oscillating
during real-time propagation, the periodic pattern in total density survived
at $t=100$, which demonstrates the dynamical stability and ensures that these super-lattice solitons can be realized experimentally.

We demonstrated spontaneous spatial order in a Rashba  SO-coupled
uniform spin-1 quasi-2D ferromagnetic $(c_{2}<0)$ BEC for large SO-coupling
strengths $\gamma $, and the formation of new types of dynamically stable
super-lattice solitons using a numerical solution of the GP equation.
The total density of these super-lattice solitons have a 2D square lattice formation. The component densities have either a stripe or a 2D square lattice density modulation.
For
small SO coupling, $(- 1,0, + 1)$ and $(0, + 1, + 2)$-type solitons
are found, which develop multi-ring structure for intermediate SO coupling
strengths. In addition, a circularly-asymmetric soliton is found to appear
as the ground state for larger SO couplings. The dynamic stability of all
these solitons was established by steady real-time propagation over a long
period of time. Super-lattice solitons can also be formed in an SO-coupled
anti-ferromagnetic $(c_{2}>0)$ BEC \cite{adhikari}, the details of which
are different from the present solitons. In the anti-ferromagnetic phase,
the circularly-asymmetric and the $(0,\pm 1,\pm 2)$-type solitons are not
possible, so that solitons of type displayed in {Figs.~\ref{fig2}}(d)-(f),
\ref{fig3}(a)-(c) and (g)-(i), and \ref{fig4}(a)-(c) will not appear. These
dynamically-stable solitons deserve further experimental and theoretical
studies. A natural extension of this investigation will be a search for a
three-dimensional super-lattice soliton in an SO-coupled spin-1 BEC.

 \section*{Acknowledgments}

S.K.A. acknowledges support by the CNPq (Brazil) grant 301324/2019-0, and by the ICTP-SAIFR-FAPESP (Brazil) grant 2016/01343-7

%\end{acknowledgments}

\end{document}